\documentclass[prb,twocolumn,aps,showpacs,superscriptaddress,epsfig,longbibliography]{revtex4-2}

\usepackage{amssymb}
\usepackage{amsbsy}
\usepackage{amsmath}
\usepackage{float}
\usepackage{epsfig}
\usepackage{graphicx}
\usepackage{subfigure}
\usepackage[dvipsnames]{xcolor}

\usepackage{graphicx}
\begin{document}
	
	\title{The effect of hyperuniform disorder on band gaps}
	
	\author{Jonas F.~Karcher}
	\affiliation{{Pennsylvania State University, Department of Physics, University Park, Pennsylvania 16802, USA}}
	
	\author{Sarang Gopalakrishnan}
	\affiliation{{Princeton University, Department of Electrical and Computer Engineering, Princeton, NJ 08544, USA}}
	
	\author{Mikael C.~Rechtsman}
	\affiliation{{Pennsylvania State University, Department of Physics, University Park, Pennsylvania 16802, USA}}

	\begin{abstract}
		The properties of semiconductors, insulators, and photonic crystals are defined by their electronic or photonic bands, and the gaps between them.  When the material is disordered, {\it Lifshitz tails} appear: these are localized states that bifurcate from the band edge and act to effectively close the band gap.  While Lifshitz tails are well understood when the disorder is spatially uncorrelated, there has been recent interest in the case of hyperuniform disorder, i.e., when the disorder fluctuations are highly correlated and approach zero at long length scales.  In this paper, we analytically solve the Lifshitz tail problem for hyperuniform systems using a path integral and instanton approach. {  We find the functional form of the density-of-states as a function of the energy difference from the band edge}.  We also examine the effect of hyperuniform disorder on the density of states of Weyl semimetals, which do not have a band gap.    
	\end{abstract}
	
	\maketitle 
	
	Hyperuniform configurations of a field are those in which long-wavelength fluctuations are parametrically suppressed relative to random configurations
	\cite{torquato2003local}.  In other words, hyperuniform configurations might look disordered at short scales but not long length scales.  Examples include periodic lattices, quasicrystals, jammed sphere packings \cite{torquato2003local}, atomic density profiles in incompressible fluids such as fractional quantum Hall systems \cite{torquato2003local}, among others. An example at molecular level is in the context of self-assembling polymers \cite{chremos2018hidden}.  Wave transport within such hyperuniform systems (be it electronic, photonic, acoustic, or otherwise) has been of interest  \cite{PhysRevResearch.4.033246,Froufe-Perez:23} because of the unconventional, spatially correlated nature of the disorder, as well as the lack of scattering at long wavelengths.  
	
	In a regular crystal, the electronic states form bands, with hard energy gaps between them where there are no states. Uncorrelated disorder smears out these hard band edges, creating states with energies inside the band gap. Lifshitz tail states are an important class of such midgap states: these states are supported on rare regions of the disorder potential. Given that crystals (which are hyperuniform) have hard band edges and random potentials do not, a natural question is how the behavior around the band edge evolves if one maintains hyperuniformity while making the system increasingly disordered at short scales. Hyperuniformity is known to affect the statistics of rare regions \cite{crowley2019quantum, shi2022many}, which host Lifshitz tail states, so one might expect an impact of the correlations on such states. 
	This question was the subject of a recent numerical study~\cite{klatt2022wave}, which posed the question of whether, and to what extent, band gaps `survive' hyperuniform disorder.  
	Ref.~\cite{klatt2022wave} employed a purely numerical approach, which is limiting because the states become too rare to sample except very close to the band edge.
	
	Here, we analytically derive an expression for the density-of-states of the Lifshitz tail in hyperuniform systems, using a path integral instanton approach of Refs.~\cite{halperin1966impurity, zittarz1966theory, yaida2016instanton}.  In this approach, we map the disordered linear system onto a nonlinear Schr\"odinger equation; the solutions of this equation are solitons whose energies then yield the density-of-states.  We use the saddle point approximation, which is valid deep within the band gap, allowing us to be certain of the nature of states in the gap.  We find that the hyperuniform correlation strongly change the functional form of the energy dependence of the density-of-states.  We compare our analytical results to numerical simulations as deeply as possible within the gap, and find strong agreement.  Moreover, we also examine the effect of hyperuniform disorder on ungapped systems, namely Weyl semimetals. Refs.~\cite{nandkishore2014rare, pixley2016rare, pixley2017single} study the WSM with uncorrelated Gaussian white noise disorder and find that vanishing density of states at the Weyl point are smeared out by randomness. Contrarily, in presence of a quasiperiodic potential the Weyl point is preserved up to a threshold strength~\cite{pixley2018weyl}. In the stealthy hyperuniform case \cite{torquato2015ensemble} we find the scaling form of the density-of-states at the degeneracy point as a function of disorder strength and the amount of constrained momenta, $\chi$. We find that at a Weyl point the leading effect of hyperuniform correlations is to change the proportionality constant in the scaling form of the DOS, whereas the presence of the states and their power law decay is robust.

	\textit{Nonlinear saddle point equations of motion:} 
	We consider a clean system described by the Hamiltonian $H_0$ (which will be either a quadratic band edge or a Weyl semimetal) in the presence of a random potential $V(x)$. The Schr\"odinger equation for an eigenstate $\psi(x)$ at energy $\epsilon$ reads:
	\begin{align}
		\left[\epsilon-H_0 - V(x)\right]\psi(x) &= 0.
	\end{align}
	We assume the potential to be drawn from a correlated Gaussian ensemble, fully determined by the first and second moment:
	\begin{align}
		\langle V(x)\rangle_V &= 0, & \langle V(x) V(x') \rangle_V &= w^2 K(x-x').
	\end{align}
	Here the average $\langle \cdot \rangle_V$ of $V(x)$ is vanishing and the second moment is given by the translationally invariant correlation function $K(x-x')$ and disorder strength $w^2$. Typically it is sufficient to assume a short range correlated or white noise kernel. In this work, we are interested in studying the impact of hyperuniform correlations $\tilde{K}(q) \sim q^\alpha$ constraining the long wavelength modes $q\sim 0$ in the Fourier transform $\tilde{K}$ of the correlation kernel. A stronger constraint happens for stealthy hyperuniform correlations, where $\tilde{K}(|q|< \kappa ) =0$ for all momenta below a cutoff momentum $\kappa$. Since we only consider square lattices here, we can measure this $\kappa$ in units of the reciprocal lattice vector $\kappa =\chi |G|$.

	Following Halperin and Lax \cite{halperin1966impurity}, we can write the leading behavior of the DOS $\rho(\epsilon)$ at energy $\epsilon$ in terms of the instanton solution of the translationally invariant nonlinear equation
	\begin{align}
		&\left[\epsilon-H_0 - V(x)\right]\psi(x) = 0, \nonumber\\
		&V(x) = \chi_0 \int d^dx'\; K(x-x')|\psi(x')|^2.\label{eq:nlse}
	\end{align}
	For definiteness of this problem, we assume the normalization convention $\int |\psi|^2=1$. At fixed $E$, the smallest threshold $\chi_0$ that permits a localized instanton $\psi$ determines the leading behavior of the DOS (see discussion in Sec. \ref{sec:nlse} of Ref. \cite{sup_mat}):
	\begin{align}
		\ln \left(L^{d}\rho(\epsilon)\right) = - \dfrac{\chi_0^{2}}{w^2}\int d^dx\;d^dx'\; K(x-x')|\psi(x)|^2|\psi(x')|^2. \label{eq:scal}
	\end{align}
	Delocalized solutions $\psi$ or strongly nonlocal kernels $K$ cause the RHS to scale with the system size and lead to a vanishing contribution to the DOS.
	
	Fluctuations around this solution give a pre-exponential correction \cite{garcia2024fluctuation}, and it is beyond the scope of this work to determine its exact form. 
	
	\textit{Quadratic band edges:} Near the band edge, we can approximate the dispersion as parabolic (around a minimum or maximum this is a good approximation) and for simplicity we assume that the mass tensor is isotropic, so we have a scalar mass $m$. We work out the details of the calculation in the supplemental material \cite{sup_mat}. In order to find the scaling of the DOS \eqref{eq:scal}, we make a rotationally symmetric ansatz $\psi(x) = (v_0/\chi_0)^{-1/2}f(\xi_\epsilon^{-1}|x|)$ to solve the NLSE \eqref{eq:nlse} with $f$ decaying exponentially and the localization length $\xi_\epsilon = 1/\sqrt{-2m \epsilon}$ given by the distance $\epsilon$ to the band edge. Here $v_0$ is a free parameter and the threshold $\chi_0(v_0)$ is fixed by normalization of $\psi$.
	
	In the following lines we give an intuition for what happens in the hyperuniformly correlated case. For small localization lengths  $\xi_\epsilon\sim a$, we basically integrate over all $x'$ of $K(\xi_\epsilon|x-x'|) |f(x')|^2$ with the same weight and obtain essentially $K(\xi_\epsilon|x|)||f||^2$ as effective potential in the differential equation \eqref{eq:nlse}. Since for both the correlated and the uncorrelated case, the kernel is decaying for large $|x|$, we expect the correlations not to have a strong effect. The more interesting case is the limit of localization lengths $\xi_\epsilon\gg a$ large compared to the lattice spacing. Then the soliton has support mostly on the small wavelengths $k$, and we can approximate the action of the hyperuniform kernel $K$ with its long wavelength limit:
	\begin{align}
		V(r) = \frac{v_0}{\epsilon} \sum_k a^\alpha|k|^\alpha \widetilde{|f|^2}(k) = \frac{v_0}{\epsilon} a^\alpha|\nabla|^\alpha |f(r)|^2
	\end{align}
	Therefore, we can approximate the kernel as a (nonlocal) fractional derivative of the delta function. This means that we expect the following effective potential $V(r) \sim \frac{v_0}{\epsilon}  (a/\xi_\epsilon)^\alpha |f(r)|^2$. We note that compared to the uncorrelated case $V(r)$ is weakened by a factor of $(a/\xi_\epsilon)^\alpha$, which affects the asymptotic scaling of the DOS strongly. The choice $v_0 = E (\xi_\epsilon/a)^\alpha$ leads to the same nonlinear equation for $f$ irrespective of the presence of correlations (which we show in Sec.~\ref{sec:band_sol} of Ref.~\cite{sup_mat}).
	
	Putting our solution back into Eq. \eqref{eq:scal}, this leaves us with the DOS
	\begin{align}
		\rho(\epsilon) \sim \exp \left[ - C_\alpha\frac{\epsilon^2}{w^2}(\xi_\epsilon/a)^{d + \alpha} \right] \label{eq:bdos},
	\end{align}
	valid in the regime of large localization lengths $\xi_\epsilon\gg a$. Here $C_\alpha$ is a numerical constant. For $\epsilon$ close to the band edge, at some point $\xi_\epsilon$ hits the lattice scale $a$ and the correlations cease to matter. In this regime it is natural to expect the hyperuniform DOS to match with the uncorrelated disorder DOS. We convince ourselves that this indeed holds with a numerical analysis of the lattice NLSE.

	\begin{figure}[h]
		\includegraphics[width=.85\linewidth]{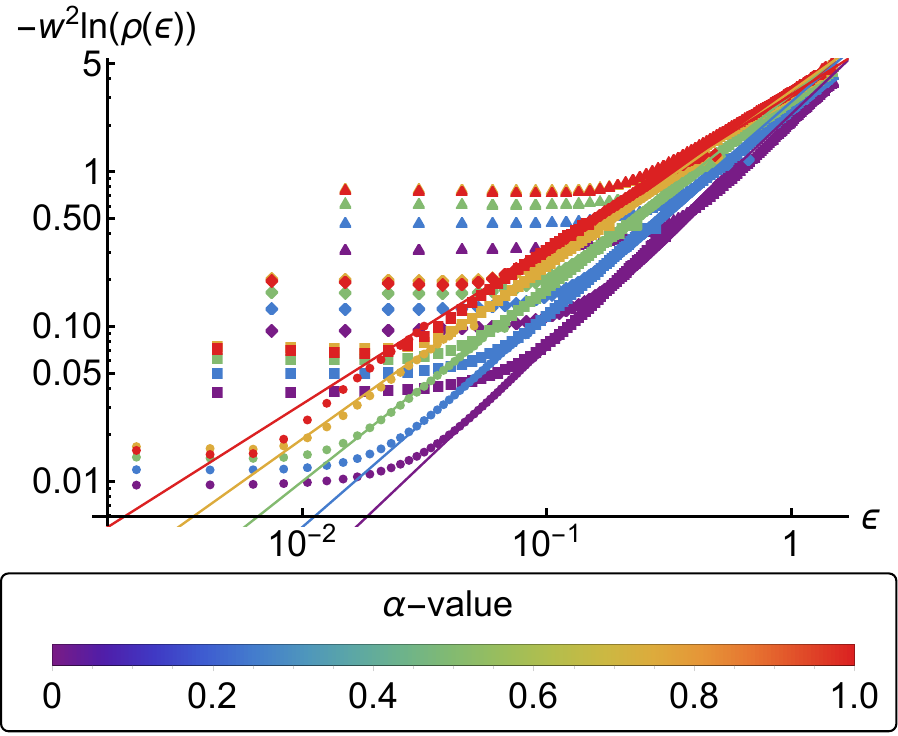}
		\caption { {\bf Numerics vs scaling for 1D disordered band edge} 
			$w^{2}\ln(\rho(\epsilon))$ is shown for four different disorder strengths $w=0.07,0.15, 0.3, 0.7$ (different markers: circle, square, diamond, triangle) and six values of $\alpha= 0.0, 0.2, 0.4, 0.6, 0.8, 1.0$ (colors). We collapse by the expected scaling~\eqref{eq:scal} for the DOS. The scaling asymptotics for different degrees of correlation $\alpha$ are shown as solid lines.
		}
		\label{fig:band_edge1d}
	\end{figure}
	
	\textit{Band edge numerics:} 
	We consider the tight binding Hamiltonian 
	\begin{align}
		H &= \sum_{\langle ij \rangle} t \;c_i^\dagger c_{j} + \sum_i V_i \; c_i^\dagger c_{i}
	\end{align}
	on square lattices in $d$ dimensions. Our systems have linear size $L$ and periodic boundary conditions. We fix the coefficient $t=1$ of the nearest neighbor hopping and disorder enters via the site dependent potential $V_i$. 
	
	We choose the family 
	\begin{align}
		K(j, l) &= \sum_{k\in BZ} \left(\sum_{r=1}^d |\sin(k_ra/2)|^2 \right)^{\alpha/2} e^{ik (j-l)}
	\end{align}
	of hyperuniform kernels on the lattice. The sum runs over all momenta $k$ in the Brillouin zone (BZ). By construction $\tilde{K}(k)\sim |k|^{\alpha}$ for long wavelengths $ka\ll 1$.

	We complement this NLSE analysis with an exact determination of the DOS in the case of disordered linear chains. We study chains of length $L=10^9$ ensemble averaged over $10^3$ configurations. The correlations are generated from independently Gaussian distributed random numbers $\xi_k$ assigned to each lattice momentum $k$ with standard deviation $W$ and mean zero with the Fourier filtering method \cite{PhysRevE.53.5445}:
	\begin{align}
		V_j&= \sum_k \tilde{V}_k e^{ik j}, & \tilde{V}_k &= |\sin(k a/2)|^{\alpha/2}\xi_k,
	\end{align}
	which can be implemented efficiently numerically using fast Fourier transforms. 
	
	In Fig.~\ref{fig:band_edge1d}, we show numerical data for the DOS of a strictly 1D chain using the Sylvester inerta method \cite{PhysRevB.18.569} for four different disorder strengths $w=0.07,\ldots, 0.7$ (different markers) collapsed by the expected scaling~\eqref{eq:scal} for the DOS. In the logarithmic axes $\ln\rho$ satisifies the expected scaling over almost two decades for different degrees of correlation, meaning that $\rho$ is controlled by our theory over 100 orders of magnitude. This data clearly shows that the correlations drastically decrease the occurence of states with large localization lengths.

	We have shown that the NLSE, scaling and brute force numerics agree remarkably well in $d=1$. We expect this to persist to $d=2,3$, where brute force numerics are computationally too demanding. Our finding is that the correlations drastically affect the scaling form  \eqref{eq:scal} for the DOS strongly suppressing the occurace of states with large localization lengths. In Sec.~\ref{sec:band} of Ref. \cite{sup_mat}, we show NLSE numerics for $d=2,3$ agreeing with the analytical result.

	\textit{Weyl semimetals:}
	In presence of disorder, the  low energy Hamiltonian near the Weyl point 
	\begin{align}
		H &=  \int d^3 {\bf r} \;\psi({\bf r})^\dagger\left[v \sum_{\mu=x,y,z}  \sigma_\mu \partial_\mu +V({\bf r})\right]\psi({\bf r}) 
	\end{align}
	contains only the velocity $v$ and the variance $W$ of the Gaussian white noise potential $V$ as parameters.
	
	As originally derived in Ref.~\cite{nandkishore2014rare}, this leads to the following scaling form of the DOS at the Weyl point
	\begin{align}
		-\ln \rho(\epsilon=0) &\sim \dfrac{v^2}{\xi^2 W^2}, \label{eq:wdos}
	\end{align}
	in the presence of Gaussian short range correlated disorder with correlation length $\xi$ (this is a technical assumption for the continuum model). At this point a comment is in order: in the WSM case, Ref.~\cite{buchhold2018vanishing} claims an exponential suppression of the DOS due to these fluctuations. In order to obtain this result, the assumption of exactly linear dispersion into the ultraviolet and a single cone is needed. In any realistic condensed matter or photonic system, the first assumption is inaccurate: there will be quadratic dispersion and termination into bulk bands. Further, Ref.~\cite{pixley2023connecting} considers the case of a single cone using short range correlated disorder, finding that it does not affect the  non-vanishing DOS in presence of disorder. 
	
	\textit{Bound states in a box potential:}
	We generalize the argument in Ref.~\cite{nandkishore2014rare} to capture the impacts of hyperuniformity and stealthiness properly. The key point here is that in the WSM case, the effective potential has a vanishing integral and decays more slowly radially. For hyperuniform Fourier filtered disorder, the effective potentials (sketched in right panel of Fig.~\ref{fig:FigHBS} or Ref.~\cite{sup_mat}) becomes longer-ranged with decay exponent smaller than $1$.

	\begin{figure}[h]
		\centering
		\includegraphics[width=.99 \linewidth]{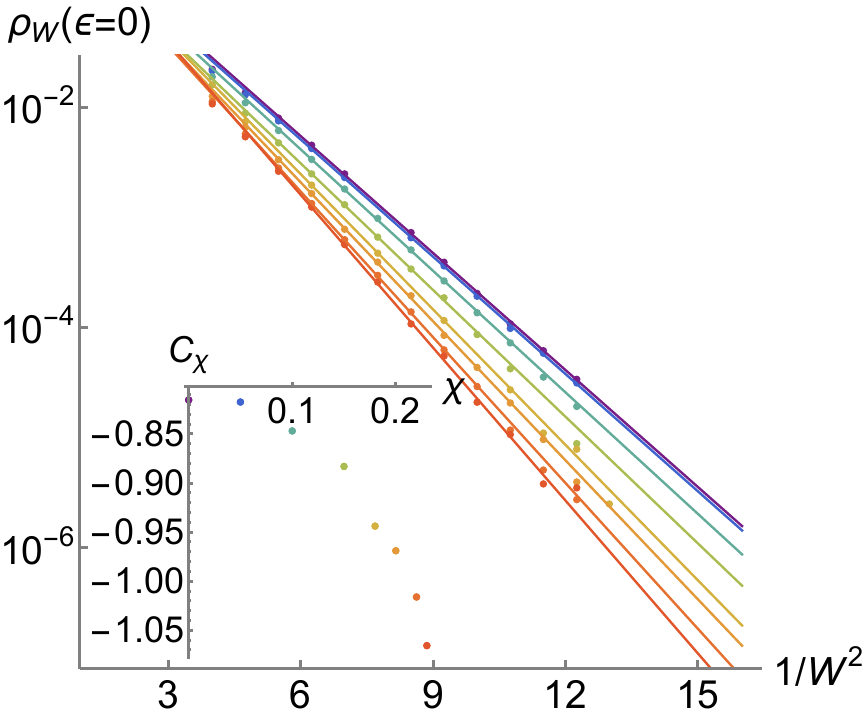}
		\caption{ {\bf Lifshitz tail states in 3D WSM with stealthy correlations} The power spectrum goes as $S(|k|<\chi)\sim 0$.  Comparison of $\rho_W(\epsilon=0)$ from data ($L=27, 37, 47, 57$, $N\leq 10^4$ configurations) to expected $W^2$ power law. For all $\chi$ (color gradient) we can observe same the $W^2$ scaling with a different proportionality constant. \textit{Inset:} Dependence of the coefficient $C_\chi$ in $\ln \rho_W(\epsilon=0) = C_\chi W^{-2} + \mathrm{const}$ on the degree of correlations $\chi$.  }
		\label{fig:d3D}
	\end{figure}

	For stealthy Fourier filtered disorder, the effective potentials (sketched in right panel of Fig.~\ref{fig:FigHBS} in Ref.~\cite{sup_mat}) not only become longer ranged with decay exponent smaller than $1$, but there are also sign oscillations with $\kappa r$. 
	
	In Sec.~\ref{sec:wsmh3d} of Ref.~\cite{sup_mat}, we show that neither the slower decay nor the oscillations spoil the presence of a localized bound state. We find that effective potentials decaying faster than $r^{-1}$ allow the same class of power law localized bound states as in the uncorrelated case. 
	
	\begin{figure}[h]
		\centering
		\includegraphics[width=\linewidth]{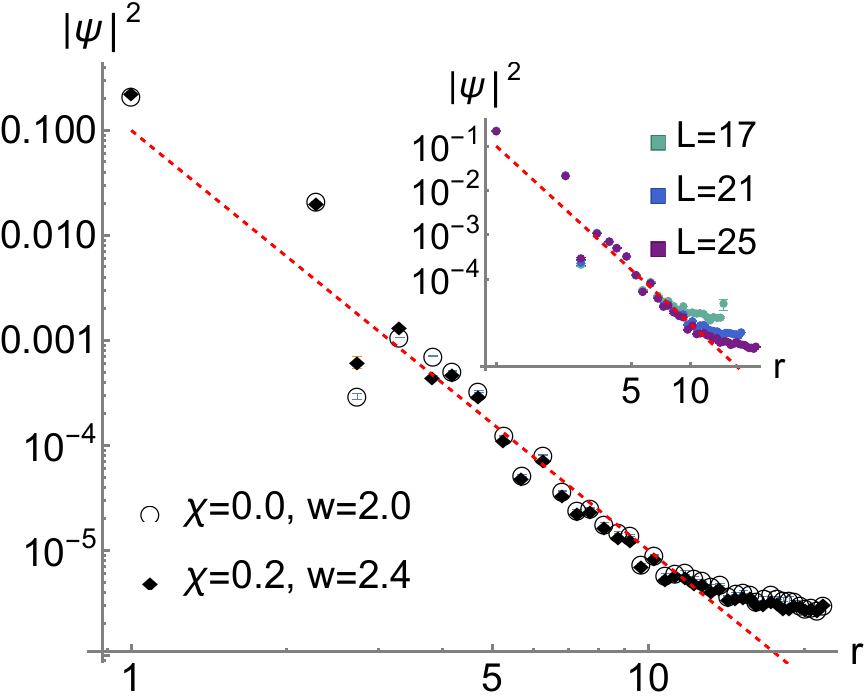}
		\caption{{\bf Radial structure of typical wavefunction $|\psi|^2(r)$} We show $L=25$ and plot the average of $|\psi|^2(r)$ (with $r$ measured from the maximum of $\psi$) for low energy wavefunctions $\psi$. Irrespective of the presence of correlations ($\chi=0$ vs $\chi=0.2$), the expected $r^{-4}$ power law is satisfied. Remarkably, the data points almost coincide in absolute value. 
			\textit{Inset:} finite size scaling for $L=17,21,25$ (colored markers). The data approaches the power law $r^{-4}$ (red dashed line) for larger $L$. }
		\label{fig:d3R}
	\end{figure}
	
	\textit{Numerics:}  Following Refs. \cite{pixley2016rare, pixley2018weyl} we study the lattice model 
	\begin{align}
		H &= \sum_{{\bf r},\mu=x,y,z} \frac12 \left(it_\mu \psi_{\bf r}^\dagger \sigma_\mu \psi_{{\bf r}+\mu} + h.c. \right) + \sum_{\bf r} \psi_{\bf r}^\dagger V({\bf r}) \psi_{\bf r} \label{eq:ham_weyl}
	\end{align}
	exhibiting a pair of Dirac cones. The spinors $\psi_{\bf r}$ have a site and spin degree of freedom. We choose isotropic nearest neighbor hopping $t_\mu = 1$. The disorder enters with the site dependent scalar potential $V({\bf r})$. We consider a system of linear size $L$ with twisted boundary conditions in all spatial directions in order to push away all plane wave eigenstates from zero energy, the leading finite-size effect, which is not relevant to the bound states \cite{pixley2016rare, pixley2018weyl}. We choose a family of stealthy kernels that does not scatter momenta outside $\chi <\pi|k|<0.25$.
	
	We make use of different techniques to numerically find the DOS of the Hamiltonian \eqref{eq:ham_weyl}. First, we use exact diagonalization (ED) with spectral folding which is efficient at higher DOS in the intermediate disorder regime. To access the low disorder regime with the most rarefied DOS, the shift invert technique is more suitable and outperforms spectral folding. In order to find the small but finite contribution of the rare tail states, ensemble averaging over $N=10^5$ configurations is crucial for small disorder.
	
	Further, we complement the ED approach with the direct determination of the DOS using the kernel polynomial method (KPM) \cite{Groth_2014}. In order to accurately determine the DOS, a large expansion order $N_e\geq 2048$ is necessary. In all our results we made sure convergence was reached by considering and comparing $N_e=1024, 2048, 4096$ (see Sec.~\ref{sec:kpm_conv} in Ref. \cite{sup_mat}).
	
	In Fig.~\ref{fig:d3D}, we analyze the disorder strength dependence of the DOS at zero energy $\rho(\epsilon=0)$ obtained with KPM and ED. The KPM data is obtained from linear system sizes $L=27,37,47, 57$ averaged over $N\leq 10^3$ configurations. The ED is obtained for $L=17,21,25$ with $N\leq 10^5$ configurations. Both datasets agree with each other and show the expected $W^{-2}$ power law \eqref{eq:wdos}. 
	
	Further, in Fig.~\ref{fig:d3R}, we show numerical data for the average of $|\psi|^2(r)$ (with $r$ measured from the maximum of $\psi$) for low energy wavefunctions $\psi$. Irrespective of the presence of correlations, the expected $r^{-4}$ power law is satisfied. Remarkably, the data points with and without correlations almost coincide in absolute value. All $\psi$ are computed with Lanczos for $10^4$ configurations of systems with linear sizes $L=17,21,25$. We verified that the expected power law is indeed approached increasing $L$. These results match with the  robustness of the bound state in the formalism described in Sec.~\ref{sec:bound} of Ref.~\cite{sup_mat}.

	\textit{Discussion:} 
	The NLSE instanton calculus is a powerful tool to gain understanding about rare region phenomena both for quadratic band edges and the WSM. Here we extended the existing formalism to capture the effect of hyperuniform correlations which drastically affect the profile of the density-of-states both in gapped and ungapped systems.  In particular, we captured the functional form of the DOS as a function of disorder and hyperuniformity parameter $\alpha$, which in turn fully describes the nature of the band gap. For the conventional band edge, we compare a scaling theory analysis to the numerical solution of that NLSE. In one dimension we can  benchmark against ED in very large systems (linear size $L=10^9$). In the case of the WSM, we use KPM and ED numerics to support the scaling theory analysis of the nonlinear Weyl equation. In both cases, we find remarkable agreement between theory and numerics persisting over many orders of magnitude of system volume.
	
	In the band edge case, the typical rare region state wavefunction appears evenly delocalized for scales smaller than the localization length. In the part of the spectrum where $\xi_\epsilon =(-2m\epsilon)^{-1/2}$ is large compared to the lattice scale $a$ hyperuniform correlations drastically alter the scaling of the DOS. For the WSM, the typical states are power law localized and do not appear delocalized at any scale (see Sec.~\ref{sec:power} of Ref.~\cite{sup_mat}), which leaves them unaffected by the correlations. We confirmed these intuitions both numerically in ED and by solving for bound states using the mapping to the NLSE.
	
	The instanton methods used in this work are best suited to the case of Gaussian disorder. For bounded disorder potentials, the nature of the tail states is different and might qualitatively modify our conclusions. We briefly discuss this for the case of quadratic band edges. Gaussian disorder leads to tail states that are tightly localized in anomalously deep potential minima. These minima do not exist for bounded potentials: instead, the tail states live in rare regions consisting of $N$ adjacent sites where the potential takes its minimum value. The kinetic energy of this type of tail state scales as $N^{-2/d}$, so the states in the extreme tail are spread out over many sites, rather than tightly localized in one. The effect of hyperuniformity---and in particular the bounded-hole property---on the existence of tail states in this context is an interesting question for future work.
	
	We expect the result above will provide a fundamental understanding of the nature of band gaps in disordered systems, including light in photonic crystals, sound in composite media, among others.

	\textit{Acknowledgement:}
	M.C.R. and J.K. acknowledge the support of the Army Research Office under the MURI program, grant number W911NF-22-2-0103.
	
	\bibliography{Correlated}

\onecolumngrid
\appendix 
\setcounter{figure}{0}
\renewcommand{\figurename}{Fig.}
\renewcommand{\thefigure}{S\arabic{figure}}

\begin{center}
	\textbf{\Large Supplemental material}
\end{center}

\section{Disorder averaged effective action formalism}
\label{sec:nlse}
We use the formalism presented in Ref. \cite{crowley2019quantum}.  The eigenstates of a clean system $\hat{H}_0$ perturbed by a disorder potential $V(x)$ satisfy the Schroedinger equation:
\begin{align}
	\left[ \hat{H}_0 + V(x) \right]\psi_n^{(V)}(x) &= \epsilon_n^{(V)}\psi_n^{(V)}(x).
\end{align}
For a fixed configuration of $V$, we can write the density of states $\rho^{(V)}(\epsilon)$ in terms of the eigenstates:
\begin{align}
	\rho^{(V)}(\epsilon) &= \dfrac{1}{L^d}\sum_{n} \delta(\epsilon- \epsilon_n^{(V)})
\end{align}
We will be interested in the average density of states $\rho(\epsilon)$:
\begin{align}
	\rho(\epsilon) &= \dfrac{1}{L^d}\int\mathcal{D}[V] \exp\left( -\dfrac{1}{2W^2}\int d^dx\;d^dx'\; V(x)K^{-1}(x,x')V(x')\right) \sum_{n} \delta(\epsilon- \epsilon_n^{(V)}),
\end{align}
where we assumed that the disorder potential is drawn from a Gaussian distribution with kernel $K$.

We can write the $\delta$-function as path integral:
\begin{align}
	\sum_{n} \delta(\epsilon- \epsilon_n^{(V)}) = \int \mathcal{D}[\psi(x),\chi(x),\Upsilon]\exp\left[i\int d^dx\; \chi^\dagger(x)\left(\epsilon -H_0 - V(x) \right)\psi(x)+i\Upsilon \int d^dx\;\psi^\dagger\psi(x) - i\Upsilon\right].
\end{align}
This gives an effective action $S_{\rm eff}[V, \psi, \chi, \upsilon]$ in the exponential.
\begin{align}
	&\rho(\epsilon) = \dfrac{1}{L^d}\int\mathcal{D}[V, \psi(x),\chi(x),\Upsilon] \exp\Big[i S_{\rm eff}[V, \psi, \chi, \upsilon]\Big],\\
	&S_{\rm eff}[V, \psi, \chi, \upsilon] = \dfrac{i}{2W^2}\int d^dx\;d^dx'\; V(x)K^{-1}(x,x')V(x') + \int d^dx\; \chi^\dagger(x)\left(\epsilon -H_0 - V(x) \right)\psi(x)+\Upsilon \int d^2x\;\psi^\dagger\psi(x) - \Upsilon.
\end{align}
We then extremize the whole action $S_{\rm eff}$ by demanding its variation to vanish in order to find the saddle point equations:
\begin{align}
	V_0(x)&=W^2 \int dx'\;K(x,x')\chi_0^\dagger(x') \psi_0(x') & \int d^2x\;\psi_0^\dagger\psi_0(x)&=1 & \Upsilon_0 &=0\nonumber\\
	0&=\left(\epsilon -H_0 - V_0(x) \right)\psi_0(x) & \chi_0^\dagger(x)\left(\epsilon +H_0 - V_0(x) \right) &=0 & \Rightarrow \chi_0 &= \lambda \psi_0^\dagger \label{eq:pot_saddle0}
\end{align}
We arrive at the effective non-linear Schroedinger equation
\begin{align}
	\left[\epsilon-H_0 - \lambda W^2\int dx'\; K(x-x')(\psi_0^{(\epsilon)})^\dagger(x')\psi_0^{(\epsilon)}(x')\right]\psi_0^{(\epsilon)}(x) &= 0.
\end{align}
for the normalized instanton wavefunction $\psi_0$.

In order to find the saddle-point contribution to the LDOS, we have to seek out the minimal $\lambda$ for which above equation has properly normalizable solutions.

This gives the leading dependence 
\begin{align}
	\rho(\epsilon) &\sim \exp\left( -\dfrac{W^2\lambda^2}{2}\int dx\;dx'\; |\psi_0^{(\epsilon)}(x)|^2 K(x,x') |\psi_0^{(\epsilon)}(x')|^2\right).
\end{align}
in terms of the solution $\psi_0^{(E)}$ of the nonlinear Schroedinger equation.
Fluctuations around the saddle point give a pre-exponential correction \cite{halperin1966impurity}.

\section{Band edges}
\label{sec:band}
In this section of the supplemental material (SM), we present the details of the continuum solution to the nonlinear Schr\"odinger equation (NLSE). 

\subsection{Solution of NLSE}
\label{sec:band_sol}
We assume a band edge in $d$ dimensions with isotropic mass $m$. This means, the NLSE takes the form:
\begin{align}
	\left[E+\dfrac{1}{2m}\Delta - \chi_0 W\int d^dx'\; K(x-x')|\psi(x')|^2\right]\psi(x) &= 0.
\end{align}
With the self-consistently defined potential $V(x) = \chi_0 W\int  d^dx'\; K(x-x')|\psi(x')|^2$, this becomes
\begin{align}
	\left[E+\dfrac{1}{2m}\Delta - V(x)\right]\psi(x) &= 0,
\end{align}
We want to find the minimum threshold $\chi_0$ for a localized eigenstate $\psi(x)$ with a hyperuniform kernel $K$.

We assume a radially symmetric $\psi(x) = \xi_E^{-d/2}f(\xi_E^{-1}|x|)$ (zero angular momentum) and use the representation $\Delta = \partial_r^2 + r^{-1}(d-1)\partial_r$ of the Laplacian. This assumption implies that the effective potential is radially symmetric $V(x) = V(|x|)$.

The radial equation (for a solution with zero angular momentum, higher angular momenta have larger thresholds $\chi_0$ and their contribution to the DOS is therefore exponentially suppressed) then is
\begin{align}
	\left[1+\xi_E^2 (\partial_r^2 + r^{-1}(d-1)\partial_r) - \frac{1}{E}V(r)\right]f(\xi_E r) &= 0, & V(x) &= \chi_0 W \xi_E^{-d}\int  d^dx'\; K(x-x')|f(\xi_E^{-1} |x'|)|^2.
\end{align}
We rescale $r$ by the localization length $\xi_E = \sqrt{-2mE}^{-1}$ (the data shown in Fig.~\ref{fig:band_edge} of the main text or Fig.~\ref{fig:band_edge3d} here shows that this is a valid approach, since this scaling indeed holds numerically):
\begin{align}
	\left[1+(\partial_r^2 + r^{-1}(d-1)\partial_r) - v(r)\right]f(r) &= 0, & v(|x|) &= \frac{1}{E}\chi_0 W \int  d^dx'\; K(\xi_E |x-x'|)|f(|x'|)|^2.
\end{align}
Neglecting the potential (appropriate very far from the origin for a decaying $V(r)$), the equation
\begin{align}
	\left[1+\partial_r^2 + r^{-1}(d-1)\partial_r\right]f(r) &= 0.
\end{align}
supports exponentially localized solutions:
\begin{align}
	f(r) = c_1r^{-\frac{d-1}{2}}e^{-r} + c_2r^{-\frac{d-1}{2}}e^{r}
\end{align}
in the limit of large $r\rightarrow \infty$. The requirement of normalizability of $f$ puts the coefficient of the exponentially growing branch to zero, $c_2=0$.

In Ref. \cite{yaida2016instanton}, an overview over the exact instanton solutions for white noise disorder is given in $d=1,2,3$. When $K$ is a nontrivial kernel, an exact analytical solution is not feasible in general. In case of hyperuniform $K$, simple scaling theory gives all the necessary information to find the form of the DOS.

\subsection{Numerical NLSE}
\label{sec:band_num3d}

\begin{figure}[h]
	\includegraphics[width=1.0\linewidth]{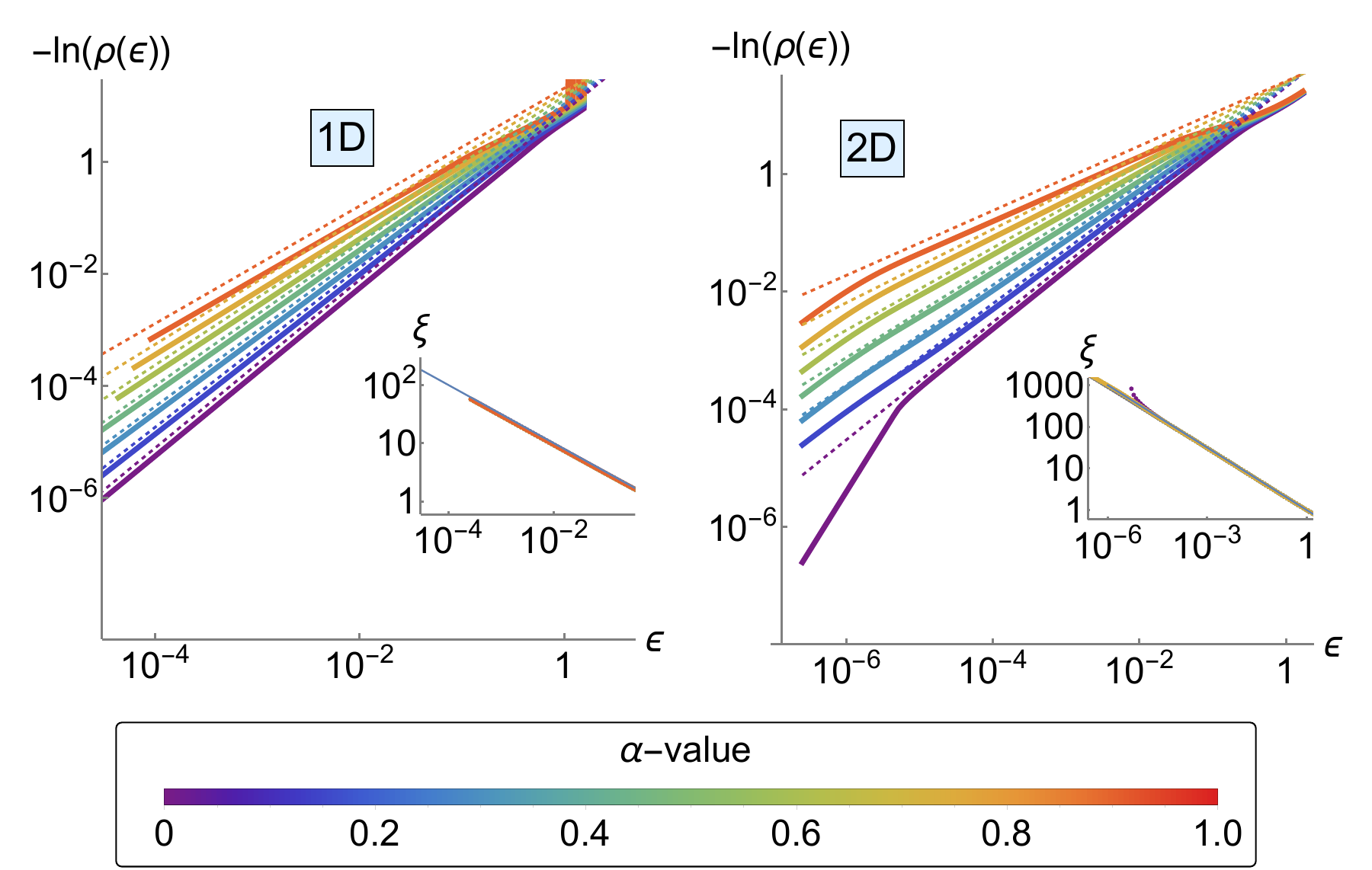}
	\caption { {\bf NLSE approach to disordered band edge} We compare the scaling theory predictions for localization length $\xi(\epsilon)$ and density of states $\rho(\epsilon)$ to the numerical solution of the NLSE for quadratic band edges in $d=1,2$ (left and right respectively) for different degrees of correlation $\alpha= 0.0, 0.2, 0.4, 0.6, 0.8, 1.0$. We show that $-\ln\rho(\epsilon)$ has a power law dependence on $\epsilon$ (solid lines). The dashed lines are the expected scaling~\eqref{eq:scal}. The data in the insets suggests that $\xi(\epsilon) = \epsilon^{-1/2}$ irrespective of correlations (data for different $\alpha$ collapses perfectly), the blue solid line is a guide to the eye.  }
	\label{fig:band_edge}
\end{figure}

In Fig.~\ref{fig:band_edge}, we compare the scaling theory predictions for localization length $\xi(\epsilon)$ and density of states $\rho(\epsilon)$ to the numerical solution of the NLSE for quadratic band edges in $d=1,2$.  In the \textit{upper panels}, we show that $-\ln\rho(\epsilon)$ has a power law dependence on $\epsilon$. The data in the insets suggests that $\xi(\epsilon) = \epsilon^{-1/2}$ irrespective of correlations. In the \textit{inset}, we show the logarithmic derivative of $-\ln\rho(\epsilon)$ to explicitly show the validity of scaling. We access average properties of systems with $10^{8}$ sites, comparison with scaling analytics (lines) holds excellent.

\begin{figure}[h]
	\includegraphics[width=\linewidth]{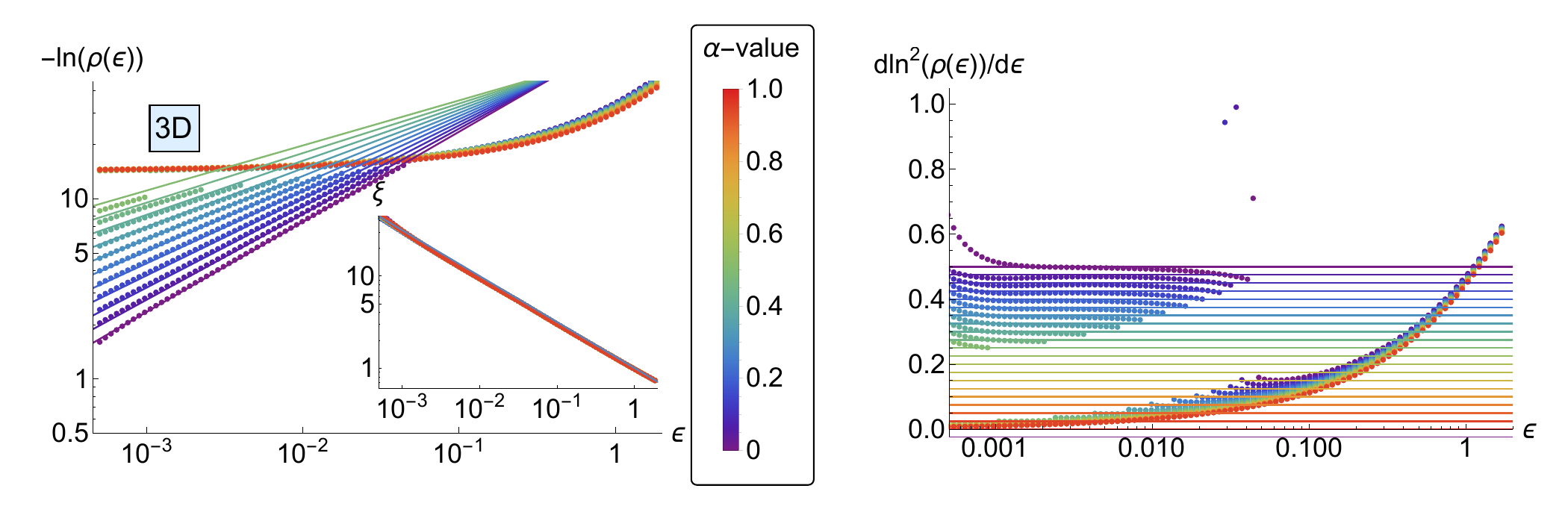}
	\caption { {\bf NLSE approach to disordered band edge} We compare the scaling theory predictions for localization length $\xi(\epsilon)$ and density of states $\rho(\epsilon)$ to the numerical solution of the NLSE for quadratic band edges in $d=3$. \textit{Left:} $-\ln\rho(\epsilon)$ has a power law dependence on $\epsilon$. The data in the insets suggests that $\xi(\epsilon) = \epsilon^{-1/2}$ irrespective of correlations. \textit{Right:} logarithmic derivative of $-\ln\rho(\epsilon)$, data approaches flat lines confirming power law behavior. }
	\label{fig:band_edge3d}
\end{figure}

In Fig.~\ref{fig:band_edge3d}, we compare the scaling theory predictions for localization length $\xi(\epsilon)$ and density of states $\rho(\epsilon)$ to the numerical solution of the NLSE for quadratic band edges in $d=3$. In the \textit{left panel}, we show that $-\ln\rho(\epsilon)$ has a power law dependence on $\epsilon$. The data in the insets suggests that $\xi(\epsilon) = \epsilon^{-1/2}$ irrespective of correlations. In the \textit{right panel}, we show the logarithmic derivative of $-\ln\rho(\epsilon)$ to explicitly show the validity of scaling. We access average properties of systems with $10^{8}$ sites, comparison with scaling analytics (lines) holds excellent.

\section{Weyl semimetal}
In this section of the SM, we present details on the analytical approach to the nonlinear Weyl equation (NLWE) and additional numerical data for the WSM in $d=2,3$.

\subsection{Solution of NLWE in 3D}
\label{sec:bound}

The NLWE with radially symmetric potential reads:
\begin{align}
	\nabla \cdot \sigma \psi  + V(r) \psi = \epsilon\psi.
\end{align}
In particular, we will be interested in the potential
\begin{align}
	V(r) &= \int d^3r\; K(r-r')\psi^\dagger\psi(r')
\end{align}
with radially symmetric kernel $K$ and self-consistently determined $\psi^\dagger\psi$.

A useful representation of the kinetic term is:
\begin{align}
	\nabla \cdot \sigma = \sigma \cdot \hat{r} (\partial_r - \dfrac{\sigma\cdot {\bf L}}{h r} )
\end{align}
It makes the symmetries of the problem visible, we need to construct our eigenstates from eigenstates of the total angular momentum ${\bf J} = \sigma+ {\bf L}$. It is 
\begin{align}
	[(\sigma \cdot \hat{r}),\sigma+ {\bf L}] = 0,
\end{align}
this means $(\sigma \cdot \hat{r})$ acts as ladder operator for $L_z$ and $\sigma_z$.

The wave function at a certain total angular momentum $j$ can be written as
\begin{align}
	\psi_{j}(r,\theta,\phi) &= 
	f(r) \Phi_{j,+}(\theta,\phi)
	+g(r) \Phi_{j,-}(\theta,\phi)
\end{align}
Here $\Phi_{j,\pm}$ carry spin $s=\pm$ and orbital angular momentum $l=j_z\mp \frac12$. They are given by \cite{nandkishore2014rare}
\begin{align}
	\Phi_{j,\pm}(\theta,\phi) &= \begin{pmatrix}
		\sqrt{\dfrac{l+\frac12\pm j_z}{2l+1}}Y_{j_z-\frac12}^{l}(\theta,\phi)\\
		\pm \sqrt{\dfrac{l+\frac12\mp j_z}{2l+1}}Y_{j_z+\frac12}^{l}(\theta,\phi)
	\end{pmatrix}.
\end{align}
Here $Y^l_m(\theta,\phi)$ are the spherical harmonics. Since we focus on the lowest angular momentum state, we do not need their explicit form in the following.

The radial equation for $f$ and $g$ is:
\begin{align}
	(\epsilon-V(r))\begin{pmatrix}
		f(r) \\
		g(r) 
	\end{pmatrix} = \partial_r\begin{pmatrix}
		g(r) \\
		-f(r) 
	\end{pmatrix} + r^{-1}\begin{pmatrix}
		(\kappa-1)g(r) \\
		(1+\kappa)f(r) 
	\end{pmatrix}, \label{eq:fg}
\end{align}
where $\kappa = j+\frac12$.

At $V=\epsilon=0$, the equations decouple and have the power-law solutions:
\begin{align}
	f(r) &= A r^{-\kappa - 1}  & g(r) &= B r^{\kappa-1}.
\end{align}
For all $j$ the function $f$ is normalizable, whereas $g$ is never normalizable, since we chose the convention $\kappa\geq 1$ to label angular momenta. We will focus on the state with $\kappa=1$ in the following, since it is natural to assume that the lowest angular momentum state has the lowest threshold value.

\subsubsection{Asymptotics for stealthy and hyperuniform correlations}
\label{sec:wsms3d}
The effective potential due to power law localized states with $f(r) \sim r^{-2}$ behaves as $V(r)\sim r^{-a}\cos(\kappa r)$ for large $r$. Hyperuniform correlations change the power $a$ away from the uncorrelated values $a=4$ and $\kappa=0$. Stealthy correlations additionally have cosine modulations of the potential.

In these lines, we show that provided that $a>1$, this allows for a self-consistent solution of the NLWE. At $\kappa=1$, $\epsilon=0$ and large $r$, we can reduce the system \eqref{eq:fg} to
\begin{align}
	-V(r)\begin{pmatrix}
		f(r) \\
		0 
	\end{pmatrix} = \partial_r\begin{pmatrix}
		g(r) \\
		-f(r) 
	\end{pmatrix} + r^{-1}\begin{pmatrix}
		0 \\
		2f(r) 
	\end{pmatrix},
\end{align}
assuming that $|g|\ll |f|$. This has the solution $f(r)\sim r^{-2}$ and $g(r)\sim r^{-a-1} \cos(\kappa r)$. Hence our assumption $|g|\ll |f|$ holds self-consistently, since $a>1$.

\subsubsection{Box potential for hyperuniform correlations}
\label{sec:wsmh3d}
Following Ref. \cite{nandkishore2014rare}, we consider the ansatz
\begin{align}
	V(r) &= \begin{cases}
		-\lambda', & r < b\\
		\lambda b^a r^{-a}, & r>b
	\end{cases}
\end{align}
for the effective potential. The hyperuniform correlations are affect the potential by choosing $\lambda$ and $\lambda'$ such to make its integral vanish $\int d^3 r V(r) = 0$ and modifying the power law decay exponent $a$.

In the inner region $r<b$, the solution of Eq. \eqref{eq:fg} with this ansatz for $V(r)$ reads:
\begin{align}
	g(r) &= A (\lambda' r)^{-1/2} J_{1/2} (\lambda' r), \nonumber\\
	f(r) &= A(\lambda' r)^{-1/2} J_{3/2} (\lambda' r)
\end{align}
with an undetermined coefficient $A$.

In the outer region $r>b$, the solution is:
\begin{align}
	f(r)&= (2-2 a)^{\frac{a+1}{2-2 a}} b^{\frac{a}{2}+\frac{1}{a-1}+1} \lambda ^{\frac{1}{a-1}+\frac{1}{2}} r^{\frac{1}{2} (-a-1)}  \nonumber\\
	&\cdot\left(c_2 \Gamma \left(\frac{3}{2}+\frac{1}{a-1}\right) J_{\frac{1}{a-1}-\frac{1}{2}}\left(-\frac{\left(\frac{b}{r}\right)^a r \lambda }{a-1}\right)-c_1 \Gamma \left(\frac{1}{2}+\frac{1}{1-a}\right) J_{\frac{1}{2}+\frac{1}{1-a}}\left(-\frac{\left(\frac{b}{r}\right)^a r \lambda }{a-1}\right)\right),\nonumber\\
	g(r)&= (2-2 a)^{\frac{a+1}{2-2 a}} b^{\frac{a}{2}+\frac{1}{a-1}+1} \lambda ^{\frac{1}{a-1}+\frac{1}{2}} r^{\frac{1}{2} (-a-1)} \nonumber\\
	&\cdot \left(c_1 \Gamma \left(\frac{1}{2}+\frac{1}{1-a}\right) J_{\frac{a+1}{2-2 a}}\left(-\frac{\left(\frac{b}{r}\right)^a r \lambda }{a-1}\right)+c_2 \Gamma \left(\frac{3}{2}+\frac{1}{a-1}\right) J_{\frac{1}{2}+\frac{1}{a-1}}\left(-\frac{\left(\frac{b}{r}\right)^a r \lambda }{a-1}\right)\right).
\end{align}
In order to eliminate coefficients, we look at the asymptotic form for $r\rightarrow \infty$:
\begin{align}
	&f(r) = 2^{\frac{1}{1-a}-\frac{7}{2}} (2-2 a)^{\frac{a+1}{2-2 a}} (a-1)^{\frac{1}{1-a}-\frac{5}{2}} b^{\frac{a}{2}+\frac{1}{a-1}+1} \lambda ^{\frac{1}{a-1}+\frac{1}{2}} r^{-5 a-2} \left(\lambda  \left(-b^a\right)\right)^{\frac{1}{a-1}-\frac{1}{2}}\nonumber\\ &\left(c_2 r^a \left(\frac{\lambda ^4 r^4 b^{4 a}}{3 a-1}-4 (a-1) \lambda ^2 r^2 (b r)^{2 a}+8 (a-1)^2 (a+1) r^{4 a}\right)\right.\nonumber\\&+\left.\frac{2^{\frac{a+1}{a-1}} (a-1)^{\frac{a+1}{a-1}} c_1 \lambda  r^3 b^a \left(\lambda  \left(-b^a\right)\right)^{-\frac{2}{a-1}} \left(\lambda ^4 r^4 b^{4 a}-4 (a-1) (5 a-7) \lambda ^2 r^2 (b r)^{2 a}+8 (a-1)^2 (3 a-5) (5 a-7) r^{4 a}\right)}{(a-3) (3 a-5) (5 a-7)}\right),\nonumber\\
	&g(r) = (2-2 a)^{\frac{a+1}{2-2 a}} b^{\frac{a}{2}+\frac{1}{a-1}+1} \lambda ^{\frac{1}{a-1}+\frac{1}{2}} r^{\frac{1}{2} (-a-1)} \nonumber\\&\left(\frac{2^{\frac{1}{a-1}-\frac{5}{2}} (a-1)^{\frac{1}{a-1}-\frac{3}{2}} c_1 r^{\frac{1}{2}-\frac{7 a}{2}} \left(\lambda  \left(-b^a\right)\right)^{\frac{a+1}{2-2 a}} \left(\lambda ^4 r^4 b^{4 a}-4 (a-1) (3 a-5) \lambda ^2 r^2 (b r)^{2 a}+8 (a-3) (a-1)^2 (3 a-5) r^{4 a}\right)}{(a-3) (3 a-5)}\right. \nonumber\\&+\left.\frac{2^{\frac{1}{1-a}-\frac{7}{2}} (a-1)^{\frac{1}{1-a}-\frac{5}{2}} c_2 r^{-\frac{9 a}{2}-\frac{1}{2}} \left(\lambda  \left(-b^a\right)\right)^{\frac{1}{a-1}+\frac{1}{2}} \left(\lambda ^4 r^4 b^{4 a}-4 (a-1) (5 a-3) \lambda ^2 r^2 (b r)^{2 a}+8 (a-1)^2 (3 a-1) (5 a-3) r^{4 a}\right)}{(3 a-1) (5 a-3)}\right).
\end{align}
We see that $g$ is not normalizable unless $c_1 = 0$.

Therefore, we get:
\begin{align}
	f(r)&= b^{\frac{a}{2}+\frac{1}{a-1}+1} \lambda ^{\frac{1}{a-1}+\frac{1}{2}} C r^{\frac{1}{2} (-a-1)} \cdot J_{\frac{1}{a-1}-\frac{1}{2}}\left(-\frac{\left(\frac{b}{r}\right)^a r \lambda }{a-1}\right),\nonumber\\
	g(r)&= b^{\frac{a}{2}+\frac{1}{a-1}+1} \lambda ^{\frac{1}{a-1}+\frac{1}{2}} C r^{\frac{1}{2} (-a-1)} \cdot  J_{\frac{1}{2}+\frac{1}{a-1}}\left(-\frac{\left(\frac{b}{r}\right)^a r \lambda }{a-1}\right).
\end{align}
with an undetermined coefficient $C$. 

Matching requires continuity (not necessarily differetiability, since Weyl is a first order differential equation) at the boundary between outer and inner region:
\begin{align}
	&A/C =  (b \lambda) ^{\frac{1}{a-1}+\frac{1}{2}}  \cdot  J_{\frac{1}{2}+\frac{1}{a-1}}\left(-\frac{b \lambda }{a-1}\right) /(\lambda' b)^{-1/2} J_{1/2} (\lambda' b),\nonumber\\
	&A/C = (b \lambda) ^{\frac{1}{a-1}+\frac{1}{2}}  \cdot J_{\frac{1}{a-1}-\frac{1}{2}}\left(-\frac{b \lambda }{a-1}\right)/ (\lambda' b)^{-1/2} J_{3/2} (\lambda' b) \label{eq:bs}
\end{align}
A caveat of this treatment is that the resulting ansatz self-consistent effective potential has a discontinuity at the boundary $b$ not present in the true  self-consistent solution. Introducing another regime could mitigate this, but then the problem is no longer tractable analytically.

\begin{figure}[h]
	\centering
	\includegraphics[width=.99 \linewidth]{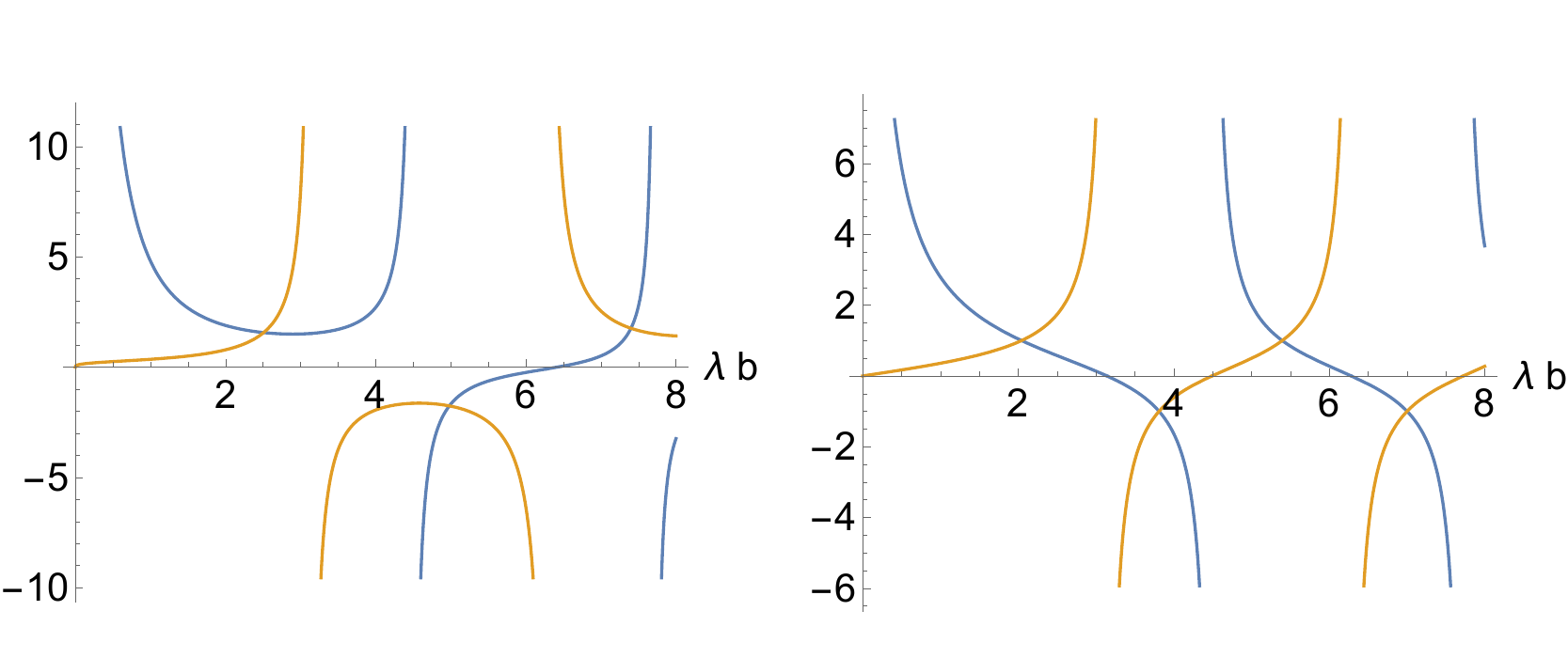}
	\caption{{\bf NLWE approach to disordered WSM} \textit{Left panel:} LHS and RHS of Eq. \ref{eq:bs} for hyperuniform $V_4$. At the intersections, bound states of the equation with hyperuniform potential occur. For any $V_\alpha$ with $\alpha>1$, the picture looks qualitatively similar.
		\textit{Right panel:} Effective potential for different correlated kernels:  uncorrelated disorder ($\alpha=0$), hyperuniform ($\alpha=1.5$) and stealthy ($\chi=0.1$ and $\chi=0.5$) with effective power laws for comparison (dashed lines). }
	\label{fig:bound}
\end{figure}

In Fig.~\ref{fig:bound}, we show the matching condition numerically for a reasonable choice of parameters and motivate the presence of a localized state at finite thresholds even in the correlated case.

\subsection{Action of hyperuniform kernels on power law localized states}
\label{sec:power}
In this section we are looking at hyperuniform correlation kernels and simple useful expressions for their long-range behavior in real space.

Let us assume the general form
\begin{align}
	K(q) = f(q)e^{-q^2 \xi^2},
\end{align}
where $f(q)$ determines the long-wavelength behavior and the exponential regulates the short wavelengths (introduction of a correlation length $\xi$).

For $f(q)=1$ this is almost ordinary white noise and we can approximate
\begin{align}
	K({\mathbf r}-{\mathbf r}') \approx \xi^d\delta({\mathbf r} - {\mathbf r}')
\end{align}
on scales $|{\mathbf r} - {\mathbf r}'| > \xi$. This approximation is valid if the kernel $K$ is applied to slow functions varying on scales larger than $\xi$.

For $f(q)=C_2 q^2$ (strong hyperuniformity), there also is a simple representation
\begin{align}
	K({\mathbf r}-{\mathbf r}') \approx C_2\xi^d\delta''({\mathbf r} - {\mathbf r}').
\end{align}
This is an almost local kernel and the approximation is again good for slow functions. Given other power-laws $f(q)=C_\alpha q^\alpha$, the kernel becomes a non-local fractional derivative.

The long distance asymptotics in that case are:
\begin{align}
	K(r) = C_1 r^\alpha \xi^{-d-2\alpha} e^{-r^2/(4\xi)} + C_2 r^{-d-\alpha\frac{d-1}{2}}
\end{align}
For even integer values of $\alpha$, the long range power law tail vanishes, i.e. $C_2=0$.

For stealthy disorder, the kernel $K$ acts as a high-pass eliminating all low frequency contributions in $f$. Therefore we cannot just assume $f$ to be slow anymore since that assumption breaks down once $K$ is applied.

The long range tail of the kernel is
\begin{align}
	K(r) = C r^{-\frac{d+1}{2}}\xi ^{-\frac{d-1}{2}}
\end{align}
in this case.

\begin{figure}[h]
	\centering
	\includegraphics[width=\linewidth]{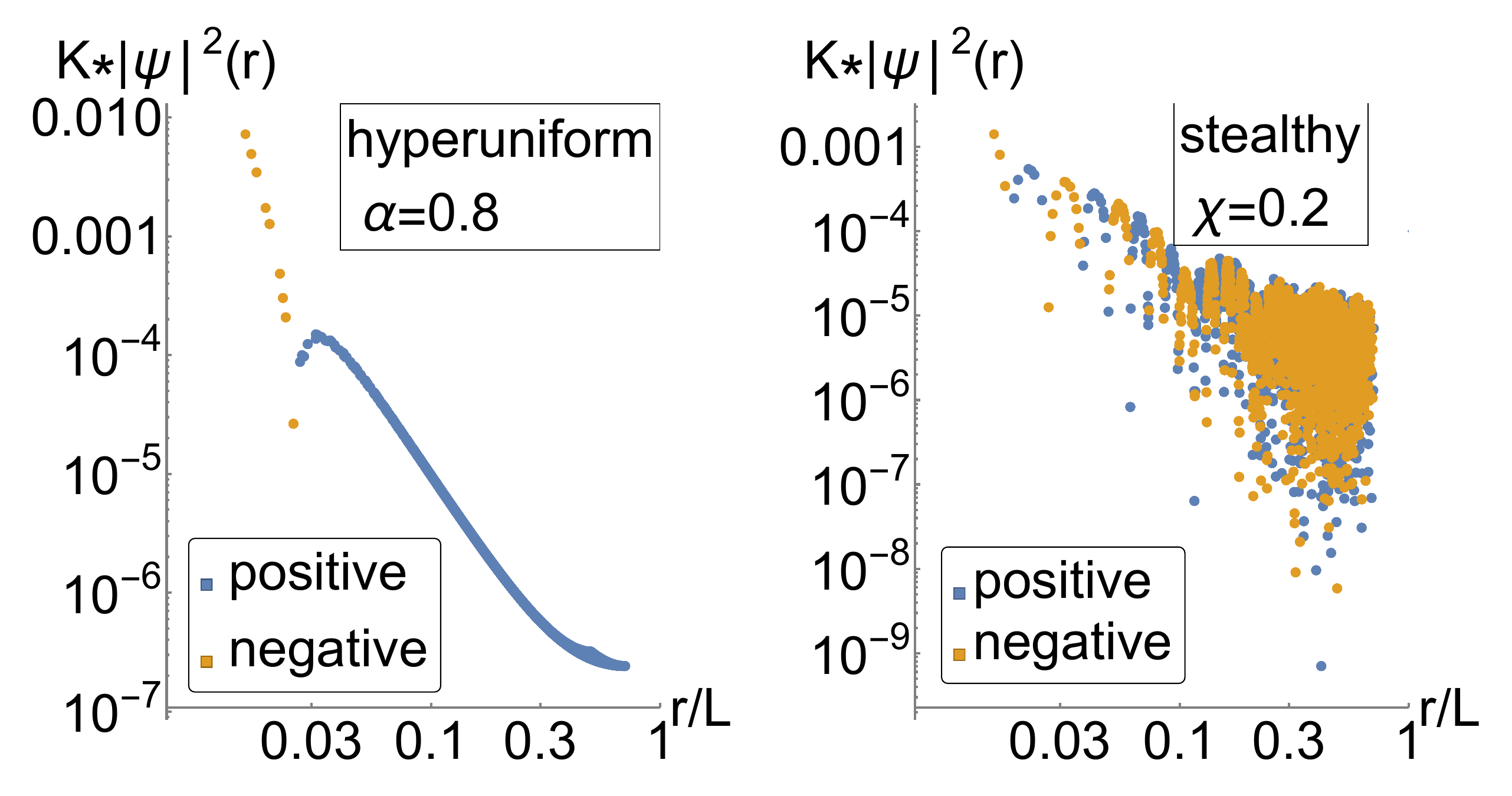}
	\caption{{\bf Effective Potential in NLWE} convolution of power law localized states with hyperuniform kernels. \textit{Left panel:} hyperuniform correlations, the approximation of a radially symmetric potential with slowly varying region near the origin and power law tail of opposed sign is well-justified.
		\textit{Right panel:} stealthy correlations, the leading part has alternating sign and decays slowly. }
	\label{fig:FigHBS}
\end{figure}

In Fig. \ref{fig:FigHBS}, we show the convolution of power law localized states with hyperuniform kernels. These appear naturally as effective potential in the NLWE. For power law hyperuniform correlations $S(k)\sim k^{\alpha}$, the approximation of a radially symmetric potential with slowly varying region near the origin and power law tail of opposed sign is well-justified. There is a sign change making the integral of the effective potential vanish. For stealthy correlations, there is more than one sign change, the leading part is oscillatory and decays more slowly.

However power law localized states $|\psi(x-x_0)|^2\sim x^{-\alpha}$ are critical in the sense that the moments of the local density of states (LDOS) $\rho(x)$ scale nontrivially:
\begin{align}
	\langle |\rho(x)|^q\rangle &\sim L^{-\tau_q}, & \tau_q &= \begin{cases}
		-(d-\alpha)q, & q<d/\alpha \\
		0, & q>d/\alpha
	\end{cases}
\end{align}
For ordinary exponentially localized states, all $\tau_q$ with $q>0$ vanish identically. In this sense the dirty Weyl semimetal could be critical at intermediate length scales even though the DOS is not vanishing. Asymptotically at large lengthscales, several power law localized states hybridize thus resolving the problem of apparent nontrivial LDOS moment scaling and criticality. Therefore it is an intriguing question what hyperuniformity does at these largest asymptotic length scales.


\subsection{Convergence of KPM expansion}
\label{sec:kpm_conv}
The KPM expansion is truncated at a fixed expansion order $N_e$. It is important to check the convergence of the numerics with respect to $N_e$.

\begin{figure*}[h!]
	\centering
	\includegraphics[width=.99 \textwidth]{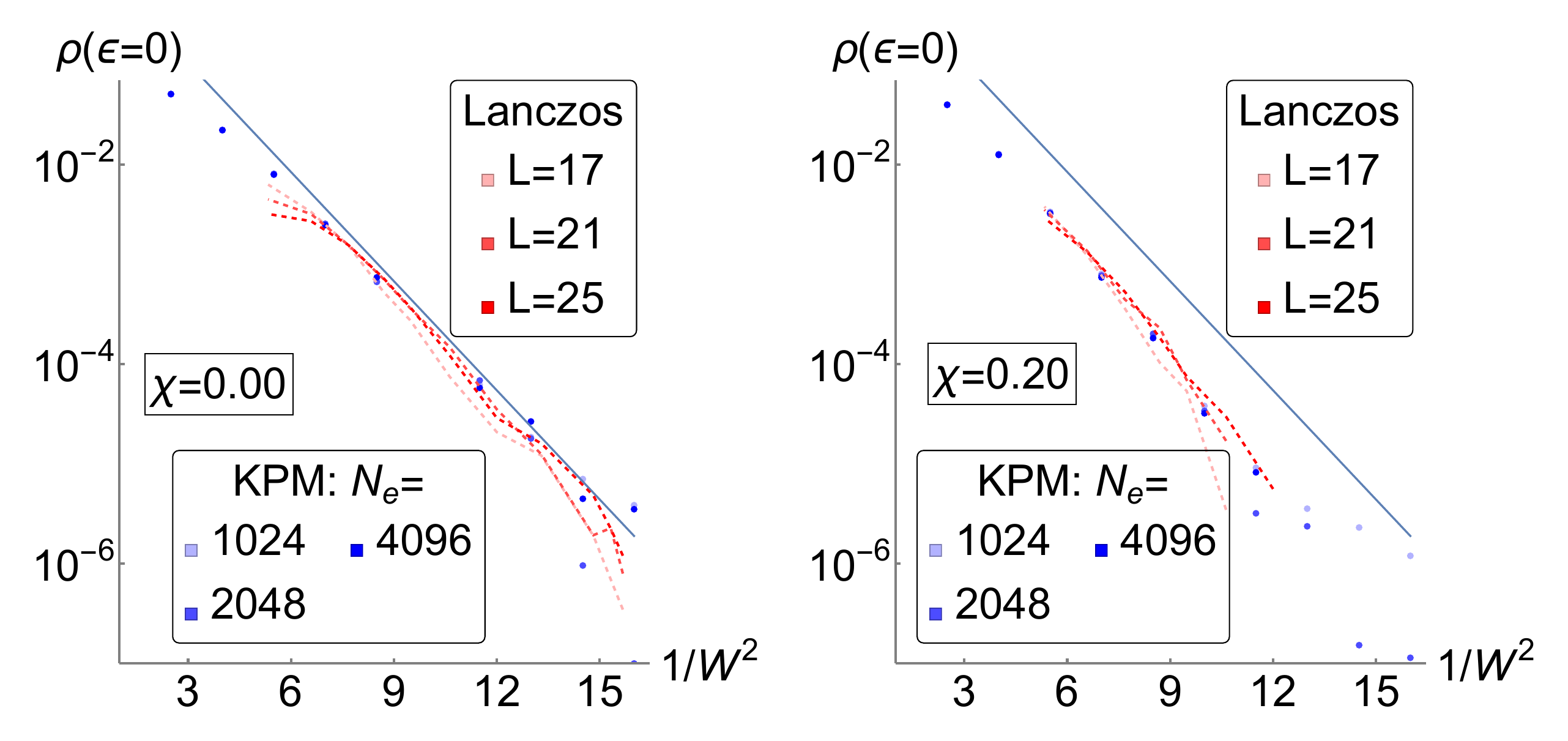}
	\caption{ {\bf Convergence of KPM expansion for 3D WSM}  Comparison of $\rho_W(\epsilon=0)$ from KPM data (blue shade dots) for $L=57$, $N= 10^3$ configurations to expected $W^2$ power law (solid blue line). In the left panel, we show uncorrelated disorder $\chi=0$, in the right panel $\chi=0.2$. In both cases, the KPM is stable in around $N_e=2048$ considered in the main text. Only for the smallest DOS $(\sim 10^{-6})$, there is weak dependence on $N_e$. }
	\label{fig:kpm_conv}
\end{figure*}

In Fig.~\ref{fig:kpm_conv}, Comparison of $\rho_W(\epsilon=0)$ from KPM data (blue shade dots) for $L=57$, $N= 10^3$ configurations to expected $W^2$ power law (solid blue line). In the left panel, we show uncorrelated disorder $\chi=0$, in the right panel $\chi=0.2$. In both cases, the KPM is stable in around $N_e=2048$ considered in the main text. Only for the smallest DOS $(\sim 10^{-6})$, there is weak dependence on $N_e$.

\begin{figure}[h]
	\centering
	\includegraphics[width=.99 \linewidth]{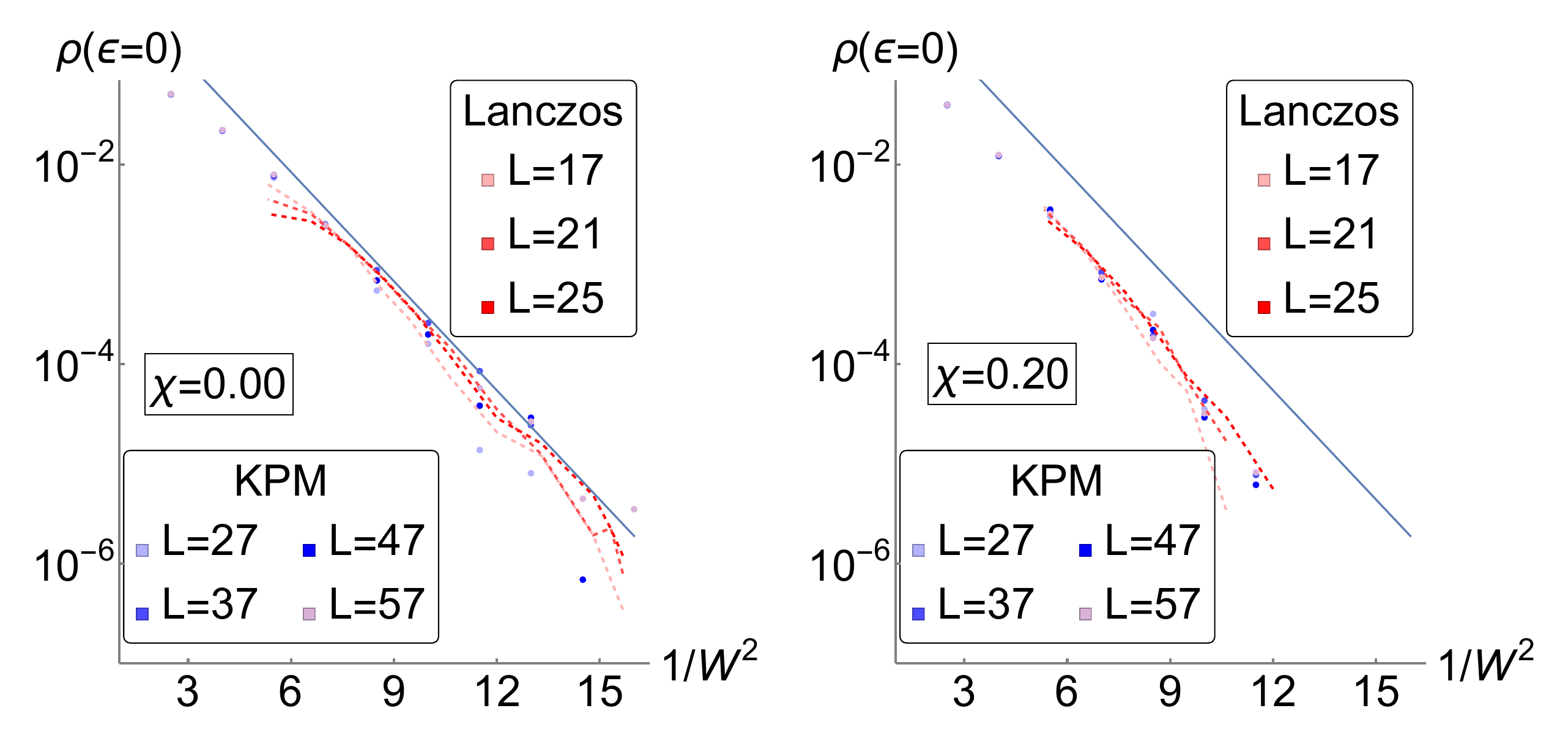}
	\caption{ {\bf Lifshitz tail states in 3D WSM with stealthy correlations} The power spectrum goes as $S(|k|<\chi)\sim 0$.  Comparison of $\rho_W(\epsilon=0)$ from data ($L=27, 37, 47, 57$, $N\leq 10^4$ configurations) to expected $W^2$ power law. For $\chi=0.2$ we can observe same the $W^2$ scaling with a different proportionality constant. This matches with the  robustness of the bound state in the formalism described in Sec. \ref{sec:bound}.  }
	\label{fig:d3d}
\end{figure}

\subsection{Correlated disorder in 2D WSM}
\label{sec:wsm2d}
Here we give a brief analysis of the 2D WSM case.
\subsubsection{Analytical treatment}
In 2D, the kinetic term in radial coordinates is:
\begin{align}
	\nabla \cdot \sigma = ( \hat{r}\cdot \sigma \partial_r  + \dfrac{1}{r} \hat{e}_\theta\cdot\sigma\partial_\theta  ) 
\end{align}
We know that $\hat{e}_\theta = 1/r(-y , x)^T$ and $L_z = h\partial_\theta$, so we can write:
\begin{align}
	= \hat{r}\cdot \sigma (  \partial_r  - \dfrac{L_z \sigma_z}{h r} ).
\end{align}
We will see that this is essentially the same form that also holds in 3D. We can write $\hat{r}\cdot \sigma = e^{i\theta} \sigma_+ + e^{-i\theta} \sigma_-$, which makes the ladder operators for angular momentum and spin visible.

To solve the radially symmetric Dirac equation, we can make the ansatz:
\begin{align}
	\psi_m(r,\theta) &= \begin{pmatrix}
		f(r) e^{im\theta}\\
		ig(r) e^{i(m-1)\theta}
	\end{pmatrix},
\end{align}
where $f$ and $g$ depend on the radial coordinate only.

The radial equation becomes:
\begin{align}
	(\epsilon-V(r))\begin{pmatrix}
		f(r) \\
		g(r) 
	\end{pmatrix} = \partial_r\begin{pmatrix}
		g(r) \\
		-f(r) 
	\end{pmatrix} + r^{-1}\begin{pmatrix}
		(m-1)g(r) \\
		mf(r) 
	\end{pmatrix},
\end{align}
which closely resembles its 3D counterpart \eqref{eq:fg}.

At $V=\epsilon=0$, the equations decouple and have the power-law solutions:
\begin{align}
	f(r) &= A r^{-m}  & g(r) &= B r^{m-1}.
\end{align}
Only for $m>1$ the function $f$ is normalizable, whereas $g$ is normalizable for $m<0$.

A state with $m=0$ leads to a log-divergent norm of $\psi$, which means that the numerics at $L=O(100)$ is likely determined by the pathological $m=0$ state that only behaves improperly asymptotically for $L\rightarrow\infty$.

\subsubsection{Numerical results}
\begin{figure*}[h!]
	\centering
	\includegraphics[width=.49 \textwidth]{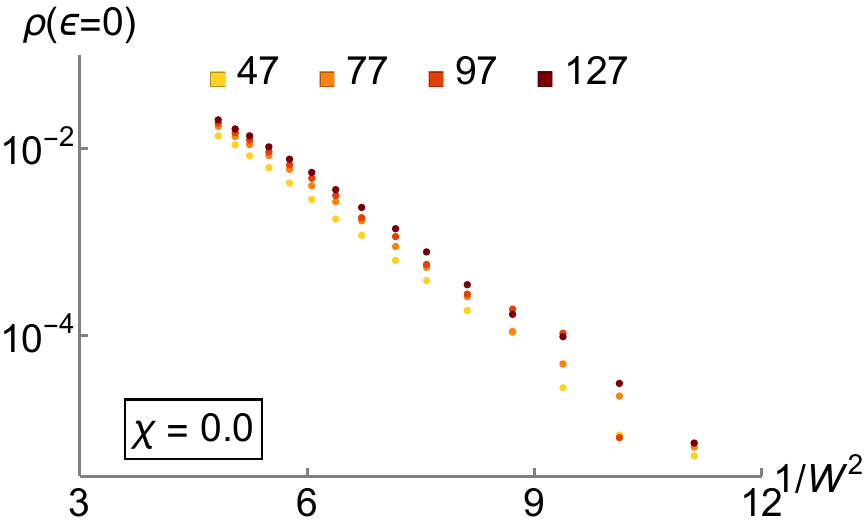}
	\includegraphics[width=.49 \textwidth]{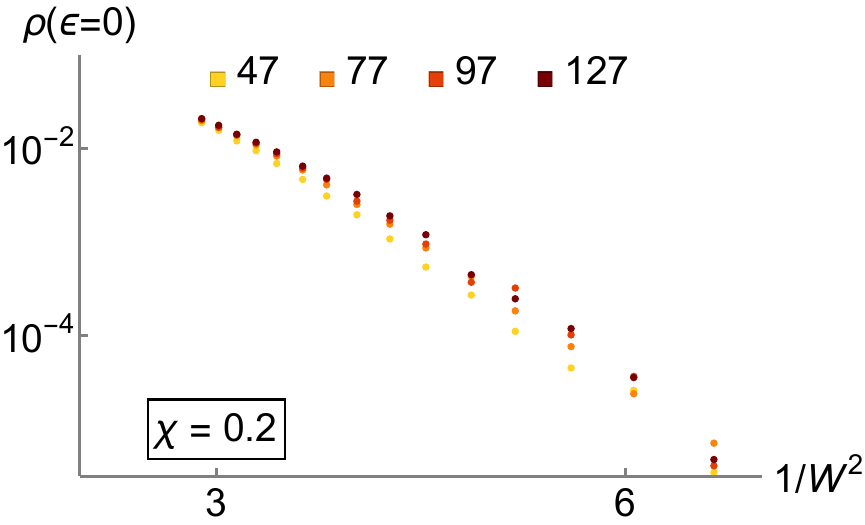}
	\includegraphics[width=.49 \textwidth]{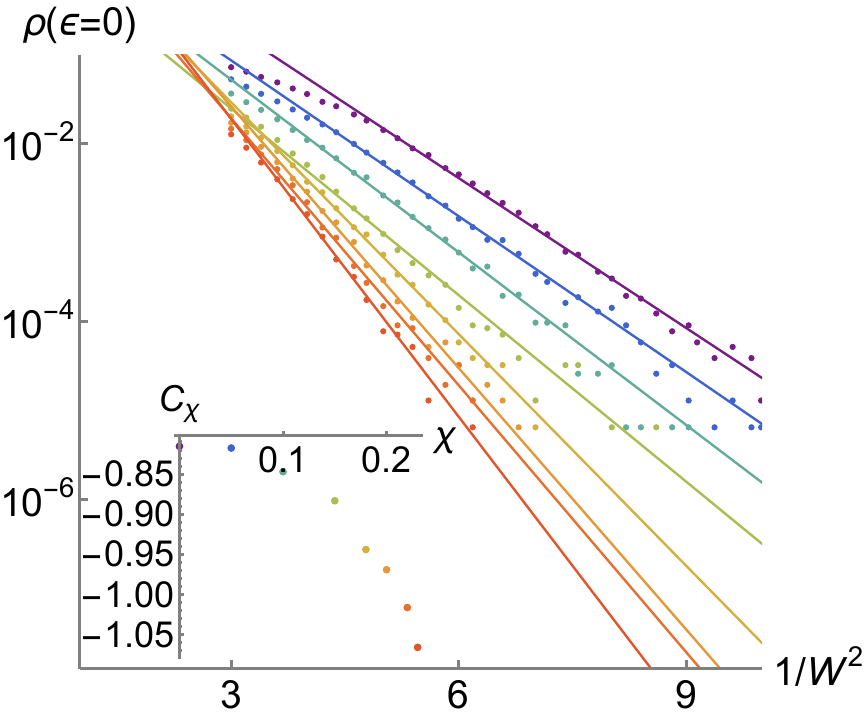}
	\includegraphics[width=.49 \textwidth]{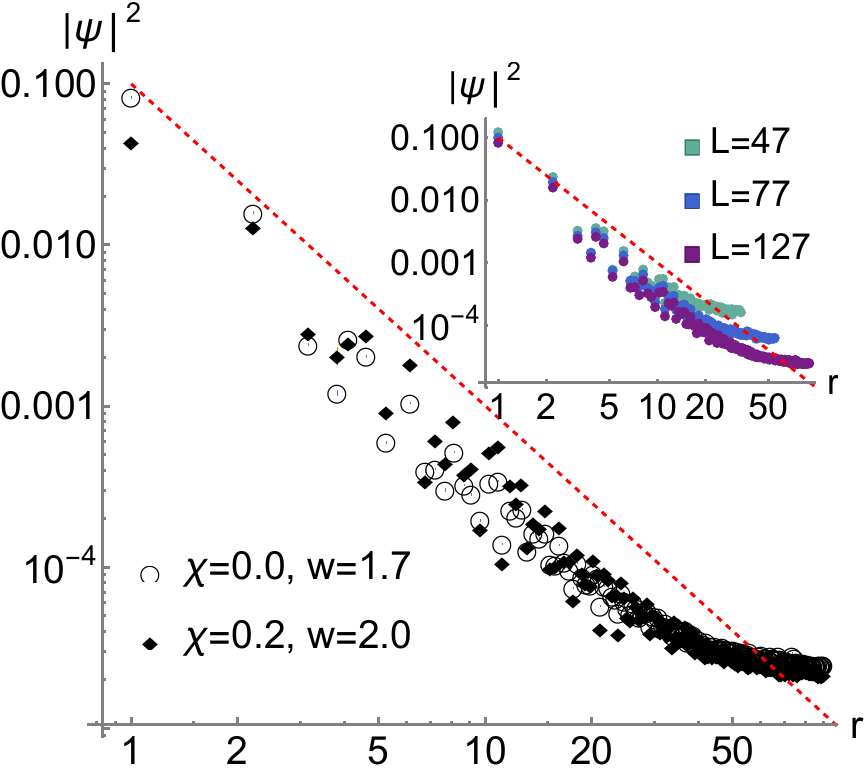}
	\caption{ {\bf Lifshitz tail states in 2D WSM with stealthy correlations} The power spectrum goes as $S(|k|<\chi)\sim 0$. \textit{Top panels:} Finite size scaling of $\rho_W(\epsilon=0)$ for $\chi=0$ (\textit{left}) and for $\chi=0$ (\textit{right}) with $L=47,\ldots,127$. \textit{Bottom left:} Comparison of $\rho_W(\epsilon=0)$ from data ($L=77$, $N= 10^5$ configurations) to the expected $W^2$ power law (rainbow colors: different degrees of correlation). \textit{Bottom right:} radial structure of the wavefunction. We see that the typical $|\psi(r)|$ goes as $r^{-2}$ (red dashed line). }
	\label{fig:d2d}
\end{figure*}

We perform an analysis fully analogous to the 3D WSM treated in the main text.
In Fig.~\ref{fig:d2d}, we show a comparison of $\rho_W(\epsilon=0)$ from data ($L=47,\ldots,127$, $N= 10^5$ configurations) to the expected $W^2$ power law. 

We find the form of this power law robust to finite size scaling of $\rho_W(\epsilon=0)$  in the range of $L=47,\ldots,127$ irrespective of correlations (\textit{upper panels} of Fig.~\ref{fig:d2d}). Our data for the radial structure of the wavefunction (\textit{lower right panel} of Fig.~\ref{fig:d2d}) further shows that  the typical $|\psi(r)|$ goes as $r^{-2}$. This shows that the DOS comes from instanton states that are not-normalizable in the thermodynamic limit $L\rightarrow\infty$ (there is a slow logarithmic divergence).

\end{document}